# Field Effect Optoelectronic Modulation of Quantum-Confined Carriers in Black Phosphorus

William S. Whitney[1,‡], Michelle C. Sherrott[2,3,‡], Deep Jariwala[2,3], Wei-Hsiang Lin[2], Hans A. Bechtel[4], George R. Rossman[5], Harry A. Atwater[2,3*]

1. Department of Physics, California Institute of Technology, Pasadena, CA 91125, USA
2. Thomas J. Watson Laboratory of Applied Physics, California Institute of Technology, Pasadena, CA 91125, USA
3. Resnick Sustainability Institute, California Institute of Technology, Pasadena, CA 91125, USA
4. Advanced Light Source, Lawrence Berkeley National Laboratory, Berkeley, CA 94720, USA
5. Division of Geological and Planetary Sciences, California Institute of Technology, Pasadena, CA 91125, USA

[‡] Equal contributors

*Corresponding author: Harry A. Atwater (haa@caltech.edu)

**Abstract:**

We report measurements of the infrared optical response of thin black phosphorus under field-effect modulation. We interpret the observed spectral changes as a combination of an ambipolar Burstein-Moss (BM) shift of the absorption edge due to band-filling under gate control, and a quantum confined Franz-Keldysh (QCFK) effect, phenomena which have been proposed theoretically to occur for black phosphorus under an applied electric field. Distinct optical responses are observed depending on the flake thickness and starting carrier concentration. Transmission extinction modulation amplitudes of more than two percent are observed, suggesting the potential for use of black phosphorus as an active material in mid-infrared optoelectronic modulator applications.

The emergence of a variety of two-dimensional materials has spurred tremendous research activity in the field of optoelectronics[1-4]. While gapless graphene can in principle exhibit an optoelectronic response at wavelengths ranging from the far infrared to the ultraviolet, its optoelectronic behavior is limited by a lack of resonant absorption and poor optical modulation in the absence of one-dimensional confinement. On the other hand, the semiconducting molybdenum- and tungsten-based transition metal dichalcogenides have shown considerable prospects for visible frequency optoelectronics. Yet while these materials promise exciting new directions for optoelectronics and nanophotonics in the visible range, they have limited response for lower energy, infrared light.

The isolation of atomically thin black phosphorus in recent years has bridged the wavelength gap between graphene and transition metal dichalcogenides, as black phosphorus is an emerging two-dimensional semiconductor material with an infrared energy gap and typical carrier mobilities between those of graphene and transition metal dichalcogenides.[5-9] Since the first isolation of black phosphorus and demonstration of a field effect device, numerous reports investigating the synthesis and optoelectronic properties of this material have emerged, appropriately summarized in recent reviews.[5, 6, 10-12] Likewise a number of reports have also appeared on the applications of black phosphorus in fast photodetectors[13], polarization sensitive detectors,[14] waveguide integrated devices[15], multispectral photodetectors[16], visible to near-infrared absorbers[17] and emitters, [18-21] heterojunction[22] and split gate p-n homojunction photovoltaics[23], gate-tunable van der Waals heterojunctions for digital logic circuits[24, 25] and gigahertz frequency transistors in analog electronics[26]. A majority of the studies on both the fundamental optical properties of black phosphorus and applications in optoelectronic devices have explored only the visible frequency range[27-30]. Therefore, little is known about the intrinsic optical response of black phosphorus in the infrared range. As a narrow band-gap semiconductor, much of the potential for black phosphorus lies in these infrared optoelectronic applications – ranging from tunable infrared emitters[31] and absorbers for waste heat management/recovery[32] to thermophotovoltaics[33] and optical modulators for telecommunications[34]. Theoretical investigations of black phosphorus have suggested novel infrared optical phenomena, such as anisotropic plasmons[35, 36], field-effect tunable exciton stark shifts[37], and strong Burstein-Moss[38] and quantum-confined Franz-Keldysh effects[39] that promise to open new directions for both fundamental nanophotonics research and applications. In this work, we report the first experimental observations of the infrared optical response of ultrathin BP samples under field effect modulation. We observe modulation of oscillations in the transmission spectra which we attribute to a combination of an ambipolar Burstein-Moss shift / Pauli-blocking effect and quantum-confined Franz-Keldysh behavior.

Measurements were performed on black phosphorous flakes that were mechanically exfoliated in a glove box onto a 285 nm $SiO_2$/Si substrate. We analyzed three flakes of 6.5 nm, 7 nm, and 14 nm thickness, determined by Atomic Force Microscopy (details are provided in the Supporting Information Fig. S6), and lateral dimensions of approximately 10 μm x 10 μm. A schematic of our experimental setup is shown in Figure 1a. Standard electron beam lithography and metal deposition methods were used to define Ni/Au electrodes to each exfoliated BP flake. The samples were then immediately coated in 90nm PMMA for protection against environmental degradation. Once encapsulated in PMMA we observe minimum degradation of our samples to ambient exposure as verified by Raman spectroscopy[40] and reported in literature precedent[41]. Transmission measurements were obtained via Fourier Transform Infrared (FTIR) Spectroscopy. All optical measurements were done in a Linkam cryo-stage at a pressure of 3 mTorr and a temperature of 80 K. First, a room-temperature gate-dependent source-drain current was measured to extract approximate carrier densities as a function of gate bias. Transmission spectra were then gathered at different gate voltages applied between the flake and lightly doped Si substrate. We note that in our setup, the silicon substrate is grounded and BP experiences the applied voltage, so the sign of the applied voltages is reversed from the more common convention. In order to probe the electric field- and charge-carrier-dependent optical properties of the BP, all spectra were normalized to the zero-bias spectrum. The measured infrared optical properties result primarily from the unique band structure of thin BP, schematically depicted in Figure 1b. Quantized inter sub-band transitions provide the primary contribution to its zero-field optical conductivity.

We first present results for the 7 nm thick BP flake, in Figure 2. An optical image is shown in Figure 2e. FTIR spectra were taken using a Thermo Electron iS50 FTIR spectrometer and Continuum microscope for which the light source is a broadband, unpolarized tungsten glo-bar. To improve signal/noise and minimize spatial drift, we surrounded the sample with a 150 nm thick gold reflector which also served as the gate electrode. The extinction modulation results are presented in Figure 2a. We observe two major features in this flake at energies of 0.5 eV (I) and 0.9 eV (II). The dip in extinction at 0.5 eV is present for both positive and negative gate voltages, as the sample is increasingly hole or electron doped, respectively. It grows in strength as the doping is further increased at larger gate-biases. The same trend is true for the feature at 0.9 eV, where a smaller peak in extinction modulation is observed for both polarities of voltage. This peak also is strengthened as the gate voltage is increased to +/- 120V. To gain insights into this behavior, we measure gate-dependent transport, using a scheme in which a positive bias induces hole-doping, and a negative bias introduces electron-doping. We observe ambipolar transport at room temperature and atmospheric conditions, as shown in Figure 2b. Similar results have been shown in the literature with on/off ratios of ~$10^4$ for flakes thinner than the one considered here, at low temperature.[22, 42] From this, the CNP is observed to be at 20 V, and, using the parallel plate model described in the Supporting Information, the unbiased, n-type carrier concentration is estimated to be $1.5·10^{12}$ cm$^{-2}$. Further discussion on the depletion length and vertical charge distribution within the flake has been provided in the Supporting Information Figure S5.

We can interpret our spectroscopic results with consideration of a Burstein-Moss shift, which is a well-known phenomenon in chemically doped narrow-band gap semiconductor materials. This effect, which changes the optical band gap of a semiconductor, results from band-filling or Pauli-blocking. As the charge carrier density is increased and the Fermi level moves into the conduction or valence band, there are fewer unoccupied electronic states available, and optical transitions to the occupied states are disallowed. This results in a decrease in the optical conductivity of the material at the energy of the transition, and is manifest in measurements as a decrease in absorption[43, 44]. Because this flake exhibits ambipolar transport behavior, we can explain both features (I) and (II) as arising from an ambipolar BM effect. At zero applied bias, the flake is very lightly doped, and all optical transitions are allowed. As a positive gate voltage is applied and the sample becomes hole doped, lower energy optical transitions become disallowed and the absorption of the flake decreases. Feature (I) corresponds to the Pauli-blocking of the $E_{11}$ intersubband transition, and feature (II) corresponds to the blocking of the $E_{22}$ intersubband transition, shown schematically in Figure 2d. For a negative gate voltage, as the sample is electron-doped and the Fermi level moves into the conduction band, the $E_{11}$ and $E_{22}$ transitions are again blocked due to band filling, resulting again in a decrease in absorption. To support this explanation, we calculate the optical conductivity for the flake, as shown in Figure 2c to identify the appropriate energies of the intersubband transitions. To do so, we use the Kubo method described by Tony Low, et al.[38] The observed transitions energies are consistent with theoretical models that predict an increase in band gap energy from the bulk 0.3 eV value as the material thickness decreases to several layers or less.[29] This deviation from the bulk band gap indicates the influence of vertical confinement of charge carriers, a feature of the two-dimensionality of the material. We note that these transition energies suggest that the true thickness of our sample is thinner than 7 nm, at approximately 4.5 nm. This apparent variation between true and observed thickness from AFM topography is a result of surface oxidation, as has been recently reported.[45] The surface oxide on our samples is expected to be between 1-2 nm on either side, which appears inevitable despite following best practices, and is stable with no measurable degradation over an ambient exposure of > 18 hrs in ambient (see Supporting Information). It is noteworthy that we observe extinction modulation at relatively high photon energies, indicative of very large charge modulation taking place in the fraction of the BP nearest to the silicon oxide interface, with an accumulation/depletion layer that decays over the remainder of the flake. This is consistent with in-depth calculations of charge screening in BP using the Thomas-Fermi model done previously, reported by Tony Low, et al.[35] We estimate this screening length to be of order 3 nm for our devices in the Supporting Information. This ambipolar, gate-modulated Burstein-Moss shift is the first observed in a two-dimensional semiconductor, to the best of our knowledge.

We next present data for a BP flake of 14 nm thickness in Figure 3. An optical image is shown in Figure 3e. Extinction measurements are again taken with an iS50 FTIR spectrometer and Continuum microscope for which the light source is a tungsten glowbar. These results are presented in Figure 3a. Four prominent features are observed to modulate under application of a gate voltage, at energies of 0.35 eV, 0.41 eV, 0.55 eV, and 0.75 eV. As in the previous sample, they grow in strength with increased magnitude of the gate voltage, regardless of polarity. To better understand this behavior, we again measure gate-dependent transport, reported in Figure 4b. We observe ambipolar transport characteristics as in the previous flake, centered about a conductance minimum at approximately 20 V. Again using a parallel-plate capacitor model, we estimate an unbiased n-type carrier density of $1.5 \cdot 10^{12}$ cm$^{-2}$ for a 20 V CNP.

We propose that the optical modulation for this sample also results from an ambipolar Burstein-Moss effect. In this case, as the Fermi energy is moved into the conduction band of the BP under negative bias, transitions become disallowed and the transmission is increased at each of the $E_{11}$ – $E_{44}$ energies. Under positive bias, as the Fermi energy is moved into the valence band, the band-filling effect of opposite charge carrier type results in negative extinction modulation peaks at the same energies of transitions $E_{11}$ – $E_{44}$. As in the previous sample, we estimate an oxide layer of 1-2 nm has grown on our BP on either surface. Based on optical conductivity calculations presented in Figure 3c, we again estimate the adjusted thickness of our flake to be less than that measured by AFM, at approximately 10 nm. We further note that for this sample, the measurement extended beyond the area of the flake, to cover the flake and an area of bare silicon oxide roughly eight times the flake are. We thus suggest that the true modulation strength of this device is of order six percent, not the 0.75 percent indicated by the modulation of the entire area.

Finally, results for the 6.5 nm thick flake are reported in Figure 4, for which an optical image is shown in Figure 4e. Unlike the previous two flakes, transmission measurements for this sample were taken using a Nicolet Magna 760 FTIR spectrometer coupled to a Nic-Plan infrared microscope on infrared Beamline 1.4.3 at the Advanced Light Source (ALS) at Lawrence Berkeley National Laboratory. This allowed us to perform measurements using a high brightness, diffraction-limited infrared beam, which is beneficial for accurately analyzing the small-area BP samples attainable by mechanical exfoliation. In contrast to the previous measurements, the incident light was elliptically polarized due to the synchrotron source, with an intensity ratio of two to one. The major axis and details of the polarization state are indicated and discussed in Supporting Information Figure S3.

Figure 4a shows the primary result of this experiment, which is the modulated extinction of the sample at different voltages, normalized to the zero-bias extinction spectrum. Three prominent features are observed in these spectra. First, under negative applied bias (i.e.: when the sample is being depleted of holes), a negative peak (I) appears in transmission near 0.45 eV, which grows in amplitude and broadens to lower energies as the magnitude of the bias increases. Second, under positive applied bias (i.e.: when the sample is being increasingly hole-doped), a positive peak (II) appears in transmittance near 0.5-0.7 eV. Lastly, these two effects, which we propose to depend on the Fermi level, are superimposed with an oscillatory feature (III) that varies with the magnitude of the applied field, but not its polarity, and which is most clearly visible in the negative bias spectra in the 0.5 - 0.7 eV range.

To better understand these results, transport measurements were again taken at room temperature under ambient conditions, as shown in Figure 4b. The gate dependence of the conductance indicates that, unlike the previous samples, this BP flake was initially heavily hole-doped, as ambipolar transport is not observed and only hole-type conduction is seen even at large negative bias. This indicates a zero-bias carrier concentration in excess of $6 \cdot 10^{12}$ cm$^{-2}$.

Due to the distinct character of each feature and their relation to the transport measurements, we can understand the overall spectral shifts as arising from a combination of a Burstein-Moss (BM) shift and a quantum confined Franz-Keldysh (QCFK) effect, both of which have been predicted theoretically for gated BP flakes of this thickness.[39] In the bulk limit, the Franz-Keldysh effect refers to electron and hole wavefunctions leaking into the band gap, as described by Airy functions. This behavior introduces oscillatory features to the interband absorption spectrum, and redshifts the band edge. In confined systems, the quantum-confined Franz-Keldysh effect similarly modulates intersubband transitions.[46] As confinement becomes stronger and excitonic effects dominate, this phenomenon eventually gives way to the quantum-confined Stark effect. Because our flake exceeds a thickness of ~4 nm, we expect excitonic effects to be weak and therefore will not focus our discussion on the quantum-confined Stark effect or a normal-to-topological phase transition in our analysis.[29, 30, 37]

We suggest that peak (I) at 0.45 eV can be described by the onset of j = 1 intersubband transitions as the material is depleted of holes at negative gate voltages and the valence band is un-filled, in agreement with our transport measurements. We further suggest that peak (II) can be described primarily by the suppression of j = 2 inter sub-band transitions as more holes are accumulated in the flake at positive gate voltages. This behavior is shown schematically in Figure 4d, and is again supported by calculations of the optical conductivity of the flake for various doping levels, shown in Figure 4e. Our experimental results correspond to modulation of the calculated intersubband transitions only in part, suggesting that a simple Burstein-Moss shift is insufficient to explain this measurement. From these results, we assign the band gap energy of our flake to be approximately 0.4 eV. Unlike our previous samples, the optical data indicates minimal oxide formation, as the $E_{11}$ and $E_{22}$ transition energies match well to theory for a 6.5 nm thick BP quantum well. Given we do not see the charge neutral point in transport, we do not assign a carrier density to this flake, but can say that with a charge neutral point of greater than -80 V, its p-type carrier density must be greater than $6 \cdot 10^{12}$ cm$^{-2}$.

We suggest that quantum-confined Franz-Keldysh effects lead to the appearance of the additional oscillatory spectral features we observe. Specifically, we point to the oscillations in the negative voltage extinction curves at energies above 0.5 eV – where Burstein-Moss considerations would predict zero modulation – and in the positive voltage extinction curves both in that same range – where Burstein-Moss behavior would predict only a single dip in extinction centered at the 0.575 transition energy – and at 0.45 eV. This oscillatory modulation increases with bias magnitude, but does not depend significantly on the sign of the bias -- behavior which is consistent with shifting of the overlap of the first and second conduction and valence sub-band wavefunctions, as described by the quantum-confined Franz-Keldysh effect. This behavior is investigated theoretically for gated BP by Charles Lin, et al.[39] In addition, under a sufficiently strong electric field, hybrid optical transitions between sub-bands of different index (eg: $E_{v1}$ to $E_{c2}$) that are nominally forbidden at zero field become allowed. In total, quantum-confined Franz-Keldysh effects in thin BP are expected to lead to behavior including redshifting of intersubband transitions, modification of intersubband selection rules (allowing hybrid transitions), or oscillatory, Airy function modulation of the absorption edge, all of which can be considered as consistent with our experimental observations. However, further theoretical work is needed to understand this effect satisfactorily; the same authors provide evidence in a more recent, experimental report that hybrid transitions may occur with zero applied field as well.[47] Interestingly, we see no evidence of a tunable plasma edge; investigations in the long-wave infrared wavelength range with larger samples would likely be needed to observe this feature.

The clear appearance of the QCFK effect in this measurement distinctly differs from our previous two samples, indicating that BP quantum wells of similar thickness may have very different optical responses. We suggest that the primary reason for this is that this flake is very heavily doped under zero bias, whereas our previous measurements were performed on nearly intrinsic flakes. In particular, in the intrinsic case, field strength and carrier concentration vary proportionally (ie: under larger bias, there is a larger carrier concentration, and vice-versa). To the contrary, in our heavily doped sample, this proportionality is absent, leading to potentially competing effects and the clear emergence of oscillatory features. It is also worth noting that, while we see no clear evidence of the QCFK effect in our first two experiments, it is possible that the large BM shift is simply dominant over the QCFK effect, making the latter effect difficult to observe, or that our increased noise prevents the effect from obviously manifesting. A complete theoretical framework that addresses the interplay between zero-bias carrier concentration and field-effect has not yet been developed, and is beyond the scope of this paper. We also note that, while we see no clear evidence of excitonic effects, and it has been suggested theoretically and experimentally that such effects should not be present in flakes of this thickness, we do not rule out the possibility that they may be influencing our results.

We note that because of the complicated polarization state of incident light from the synchrotron, and because a previous study has extensively studied this effect experimentally[48], we do not address in detail the anisotropic optical properties of BP. However, due to the primary contribution to the optical conductivity arising from the $\sigma_{xx}$ component, we argue that the only effect of elliptically polarized light is to scale the observed modulation, as discussed in the Supplement Section 1 - 3.

In conclusion, we have demonstrated experimentally that ultra-thin black phosphorus exhibits widely tunable, quantum well-like optical properties at mid-infrared wavelengths. In 7 and 14 nm, lightly doped flakes, we observe for the first time an ambipolar Burstein-Moss shift of intersubband transitions, which also varies with thickness as these transition energies are changed. In a heavily doped 6.5 nm thick BP flake, modulation of infrared transmission takes place as a result of both a Burstein-Moss shift and additional, quantum-confined Franz-Keldysh effects. While our results verify some of the recent theoretical predictions about the electro-optical effects in few-layer BP, they also report new behavior and serve as motivation to further understand the BP optical response as function of sample thickness, doping and field. Our results indicate that BP is both an interesting system for exploring the fundamental behavior of quantum-confined carriers in two-dimensional semiconductors under field-effect modulation, and a promising candidate for tunable mid-infrared optical devices.

**Methods:**

BP flakes were exfoliated in a glove box from crystals grown by HQ Graphene. After fabrication of Ni/Au (20 nm / 130 nm) electrodes by electron beam lithography and electron beam evaporation, 90 nm PMMA 950 A2 was spin-coated as an encapsulation layer. Electron beam lithography was again used to expose the contacts for wire bonding. PMMA[10] and other encapsulation layers including ALD grown dielectrics,[49, 50] polymers[41, 51], covalent surface functionalization[52] and atomically thin hexagonal boron nitride[53] have been shown in the past to successfully protect BP devices against ambient degradation.

**Acknowledgments:**

This work was supported by the U.S. Department of Energy (DOE) Office of Science, under grant DE-FG02-07ER46405. The authors gratefully acknowledge use of the facilities of beamline 1.4.3 at the Advanced Light Source which is supported by the Director, Office of Science, Office of Basic Energy Sciences, of the U.S. Department of Energy under Contract No. DE-AC02-05CH11231. M.C. Sherrott and D. Jariwala acknowledge support by the Resnick Institute and W.S. Whitney acknowledges support by the National Defense Science and Engineering Graduate Fellowship. This research used resources of the National Energy Research Scientific Computing Center, a DOE Office of Science User Facility supported by the Office of Science of the U.S. Department of Energy under Contract No. DE-AC02-05CH11231. The authors are grateful to Victor Brar for helpful discussions.

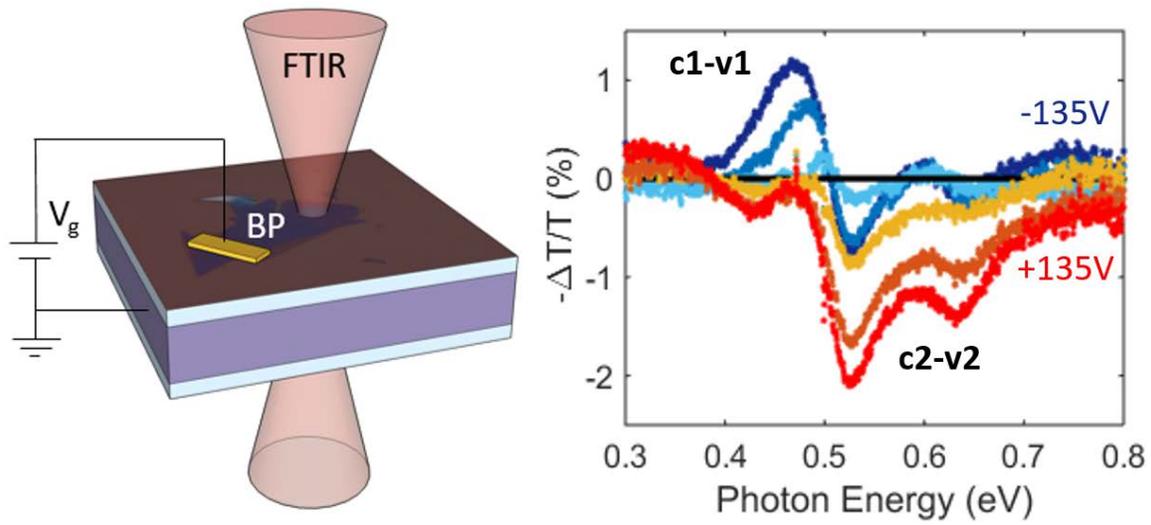

**Table of Contents Figure**

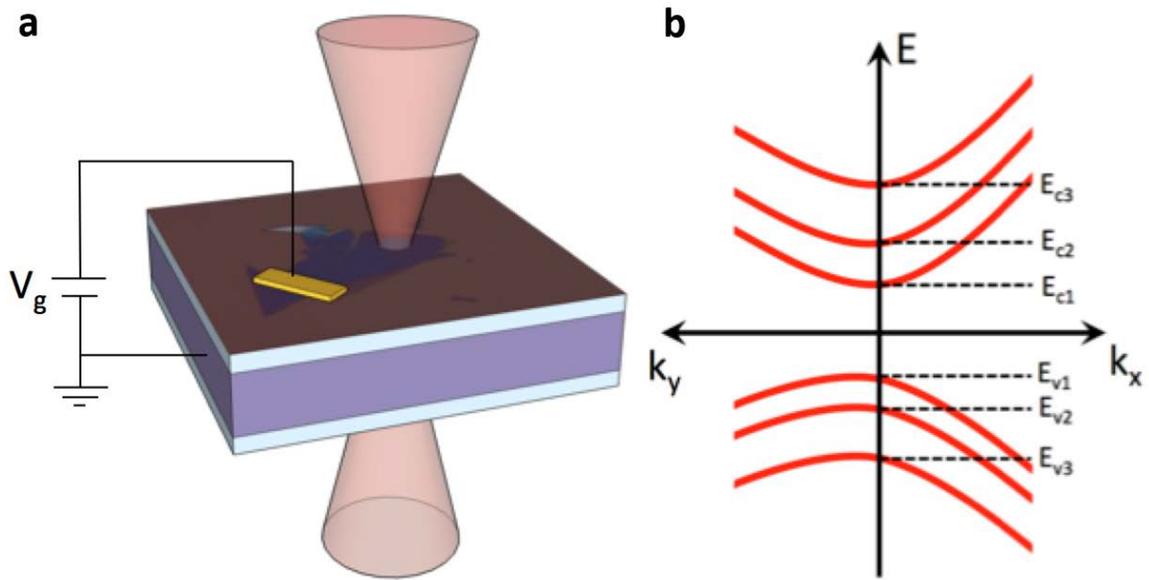

**Figure 1: a)** Schematic illustration of transmission modulation experiment. Broadband mid-IR beam is transmitted through black phosphorus sample. Variable gate voltage applied across $SiO_2$ modulates transmission extinction, **b)** Schematic band diagram of few-layer black phosphorus with subbands arising from vertical confinement

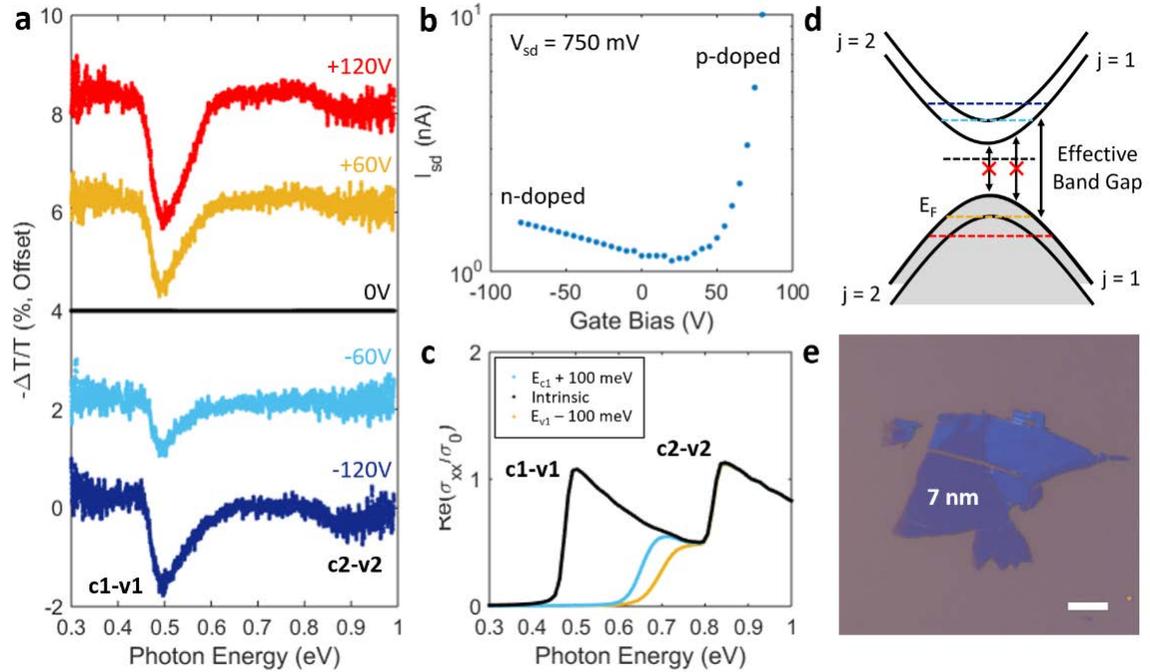

**Figure 2:** Gate modulation of lightly doped 7 nm flake. **a)** FTIR transmission extinction vs photon energy normalized to zero bias **b)** Source-drain current vs gate voltage. Ambipolar conduction is seen. **c)** Calculated optical conductivity of a 4.5 nm thick BP flake at different carrier concentrations, normalized to the universal conductivity of graphene. No field effects included. **d)** Schematic of electronic band structure and allowed interband transitions at different voltages. **e)** Optical microscope image of flake. Scale bar is 10μm.

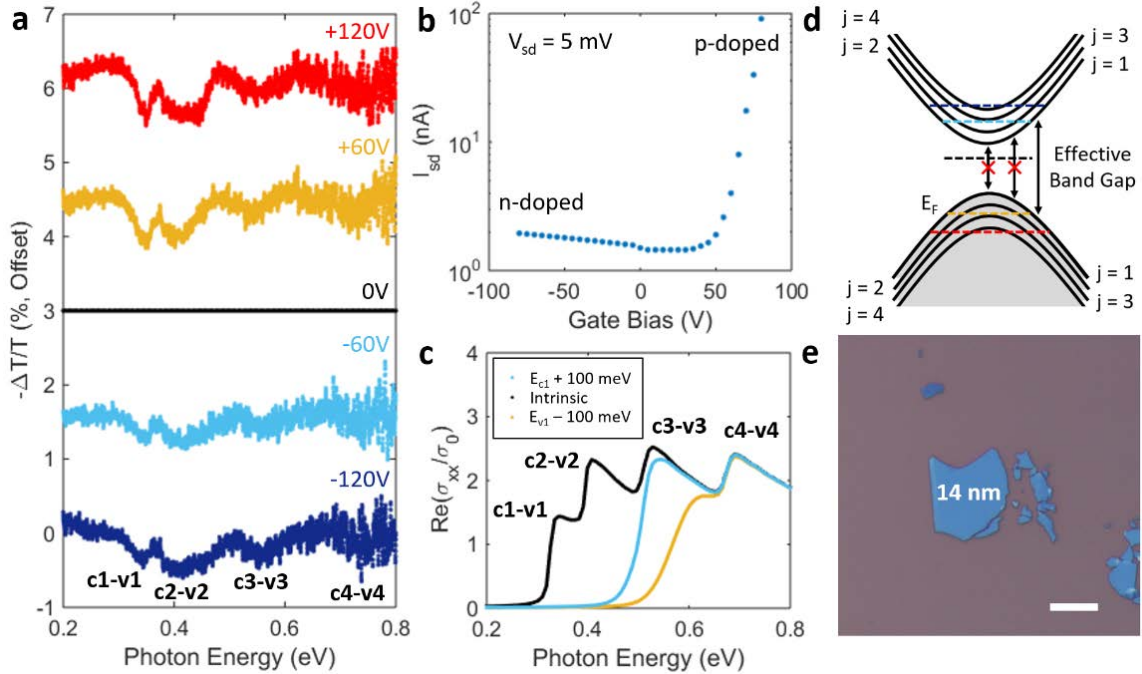

**Figure 3:** Gate modulation of lightly doped 14 nm flake. **a)** FTIR transmission extinction vs photon energy normalized to zero bias **b)** Source-drain current vs gate voltage. Ambipolar conduction is seen. **c)** Calculated optical conductivity of a 10 nm thick BP flake at different carrier concentrations, normalized to the universal conductivity of graphene. No field effects included. **d)** Schematic of electronic band structure and allowed interband transitions at different voltages. **e)** Optical microscope image of flake. Scale bar is 10 μm.

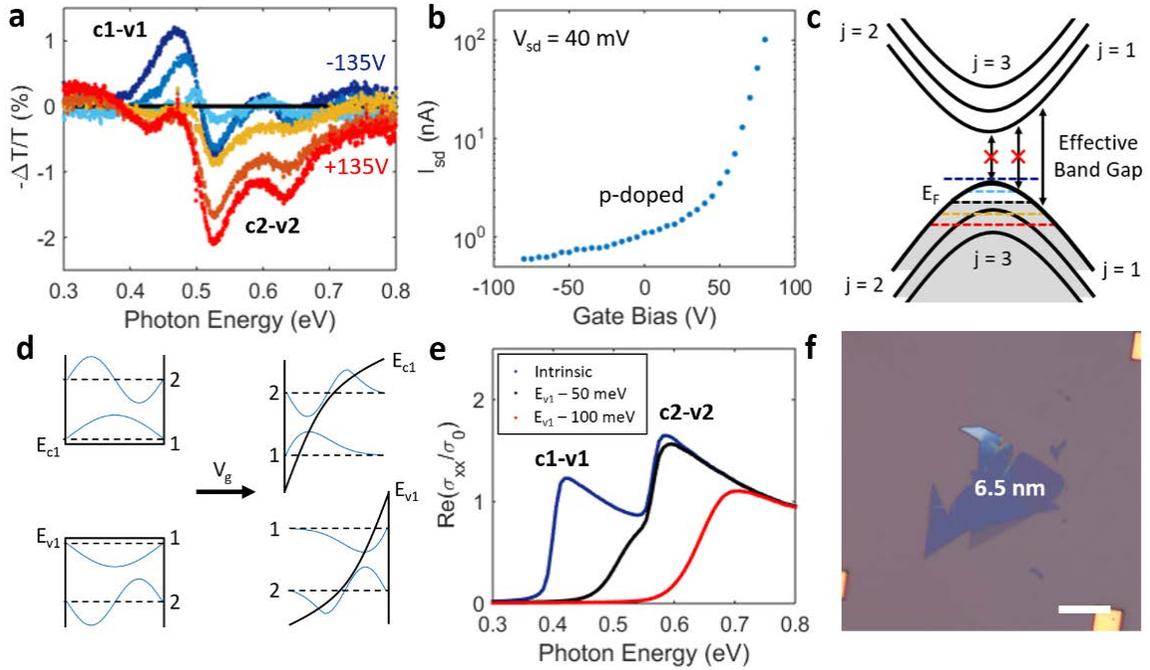

**Figure 4:** Gate modulation of a heavily doped 6.5 nm flake. **a)** FTIR transmission extinction vs photon energy normalized to zero bias **b)** Source-drain current vs gate voltage. Only hole-type conduction is seen. **c)** Schematic of electronic band structure and allowed interband transitions at different voltages. **d)** Schematic representation of quantum confined Franz-Keldysh Effect **e)** Calculated optical conductivity of a 6.5 nm thick BP flake at different carrier concentrations, normalized to the universal conductivity of graphene. No field effects included **f)** Optical microscope image of flake. Scale bar is 10 μp.

**Supporting Information for: "Field Effect Optoelectronic Modulation of Quantum-Confined Carriers in Black Phosphorus"**

William S. Whitney[1‡], Michelle C. Sherrott[2,3‡], Deep Jariwala[2,3], Wei-Hsiang Lin[2], Hans Bechtel[4], George R. Rossman[5], Harry A. Atwater[2,3*]

1. Department of Physics, California Institute of Technology, Pasadena, CA 91125, USA
2. Thomas J. Watson Laboratory of Applied Physics, California Institute of Technology, Pasadena, CA 91125, USA
3. Resnick Sustainability Institute, California Institute of Technology, Pasadena, CA 91125, USA
4. Lawrence Berkeley National Laboratories, Berkeley, CA 94720, USA
5. Division of Geological and Planetary Sciences, California Institute of Technology, Pasadena, CA 91125, USA

[‡] Equal contributors

*Corresponding author: Harry Atwater (haa@caltech.edu)

## I.  Extinction Modulation of 25 nm thick Flake

The infrared extinction results are shown in Figure S1 normalized to the zero bias extinction, and show significant modulation of a single broad feature. This feature is strongest at positive bias, and reverses sign twice: it changes polarity as the bias crosses 0 V, and again between -60 V and -120 V.

We interpret our results for the 25 nm sample as an ambipolar Burstein-Moss shift. Because this flake exhibits ambipolar transport, we can understand the primary spectral feature as resulting from three separate regimes of charge carrier modulation. At increasingly positive bias (ie: increased hole doping), Pauli blocking of optical transitions is increased, resulting in higher infrared transmission at lower photon energies. At negative bias, transmission first decreases as we deplete the sample of holes and more optical transitions are allowed, and then increases as the sample becomes electron-doped and a Burstein-Moss effect of the opposite charge carrier type is introduced. This ambipolar, gate-controlled Burstein-Moss shift is the first observed in a two-dimensional semiconductor, to the best of our knowledge. We note the presentation of this data in arbitrary units, as further investigation of modulation strength in thick BP flakes is needed to draw definitive, quantitative conclusions.

Superimposed on this large modulation are small oscillations that are most evident at high applied field – particularly +120 V. We suggest that these oscillations are related to features in the quantized intersubband transitions that occur in the BP optical conductivity, as seen from the calculation in Figure S5d. A further, larger oscillation appears in the -120 V transmittance spectrum near 0.3 eV. We speculate that this feature may result from distinctions between electron and hole-doped optical responses; however further study would be required to draw definitive conclusions about this. We further note that transport measurements for this flake were performed at 80 K, at a pressure of 3mTorr.

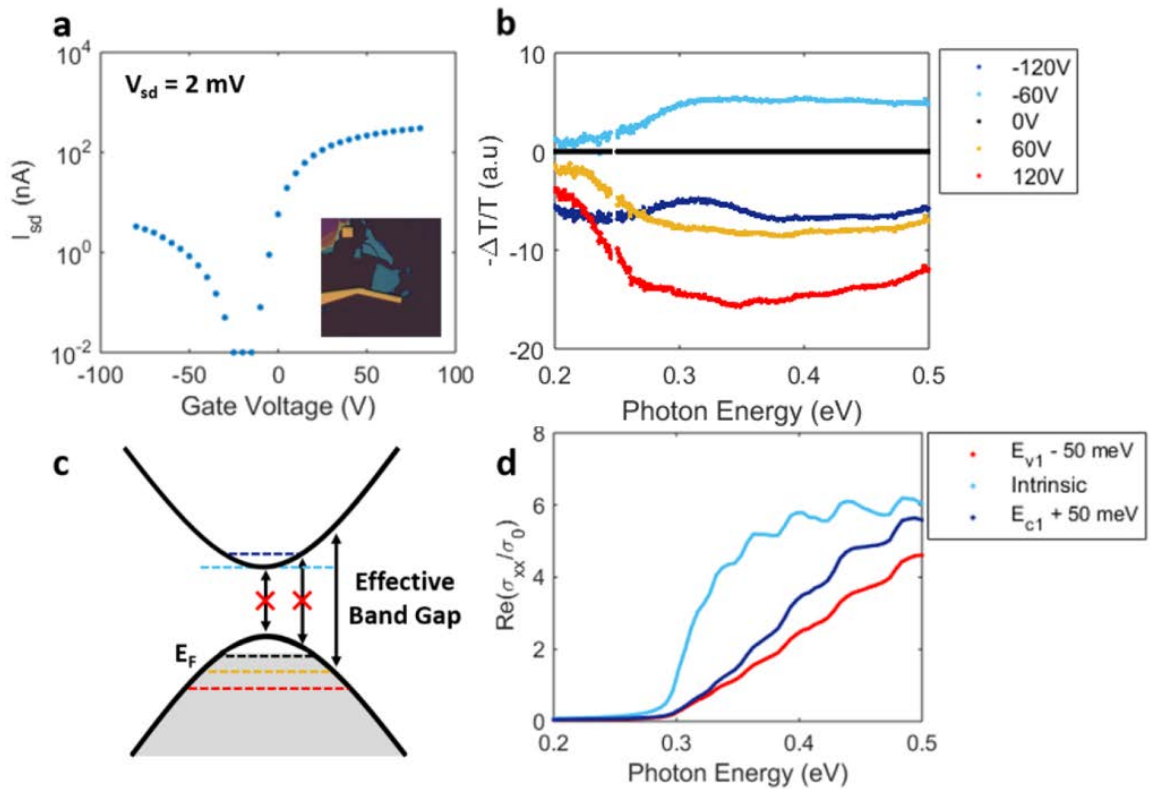

**Figure S1:** Gate modulation of 25 nm flake. **a)** Source-drain current vs gate voltage. Ambipolar conduction is seen. Inset: Optical microscope image of flake. **b)** FTIR e vs photon energy normalized to zero bias. **c)** Schematic of electronic band structure and allowed interband transitions at different voltages. **d)** Calculated optical conductivity of an intrinsic, 10 nm BP quantum well, normalized to the universal conductivity of graphene.

## II.    Crystal Lattice Structure

The x (armchair) and y (zig-zag) crystal lattice directions are determined by polarization-dependent visible reflectance measurements. At each angle of polarization an image is recorded, and pixel RGB values are sampled from both the BP flake and nearby substrate. The ratio of green channel values from flake to substrate is averaged over three sample positions, and plotted as a function of polarization angle in Figure S1. Maxima and minima in green reflectance determine the armchair and zig-zag directions, respectively.[54] This characterization was not performed for the samples measured with the internal FTIR glo-bar source, as those measurements are fundamentally unpolarized.

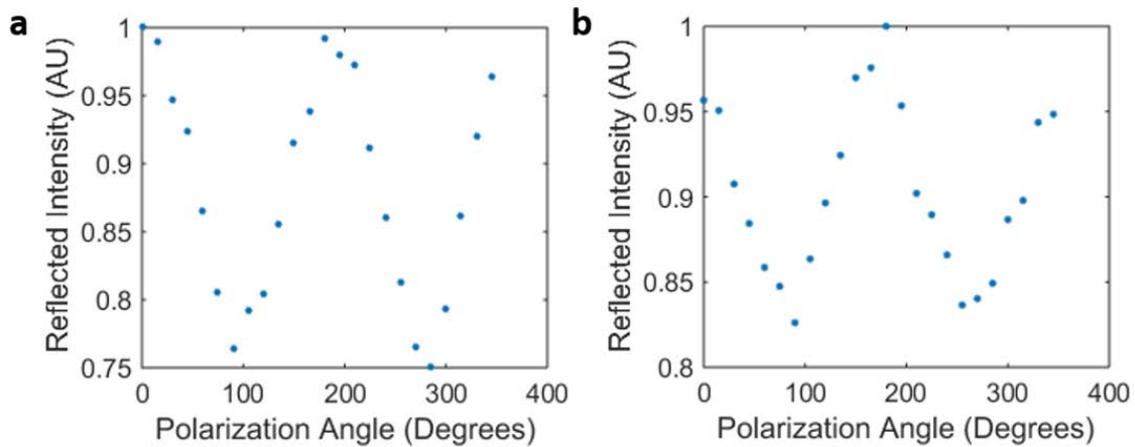

**Figure S2:** Intensity of the green channel of light reflected from BP flakes as the linear polarization of the incident light is rotated for **a)** the 6.5 nm flake and **b)** the 25 nm flake. In both cases, the polarization angle is defined as the angle between the x (armchair) crystal axis and the linear polarizer. The green component of the pixel RGB of the flakes is normalized to that of the adjacent substrate.

### III. Polarization State of Synchrotron FTIR Beam

The FTIR beam used in our final (Fig 4) measurement and Supplement S1 has an inherent elliptical polarization due to its synchrotron source. The polarization state is approximately two to one polarized along the major and minor axes of this ellipse, which are indicated in Figure S3. Due to the complicated polarization state of incident light from the synchrotron, and because a previous study has extensively investigated this effect experimentally[48], we do not study in detail the anisotropic optical properties of BP. However, since the $\sigma_{xx}$ component of the optical conductivity is one to two orders of magnitude larger than the $\sigma_{yy}$ component, plotted in Figure S4, we argue that the observed optical response derives almost entirely from light-material interactions along the armchair direction. As a result, the only effect of elliptically polarized light is to scale down the observed modulation strength. Probing devices with light of properly aligned polarization – linear along the armchair direction – would maximize this modulation strength; however, the underlying physics would be unchanged.

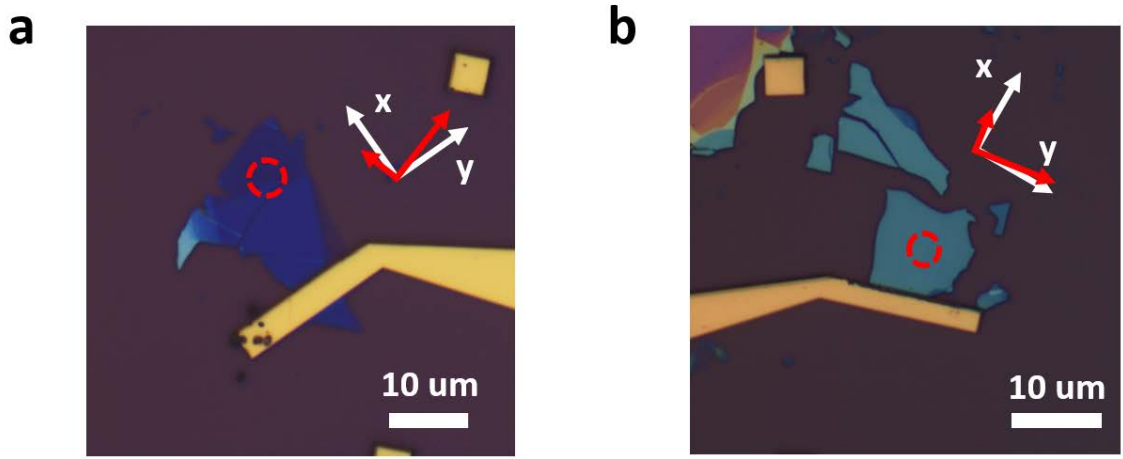

**Figure S3:** Polarization states of FTIR light. The synchrotron infrared source is inherently polarized at ALS beamline 1.4.3, with a roughly 2:1 elliptical polarization in the direction indicated here in red for **a)** the 6.5 nm flake and **b)** the 25 nm flake. Also indicated are the crystal axes, where x and y correspond to the armchair and zig-zag lattice directions, respectively, and the measurement site, indicated by a red, dashed circle.

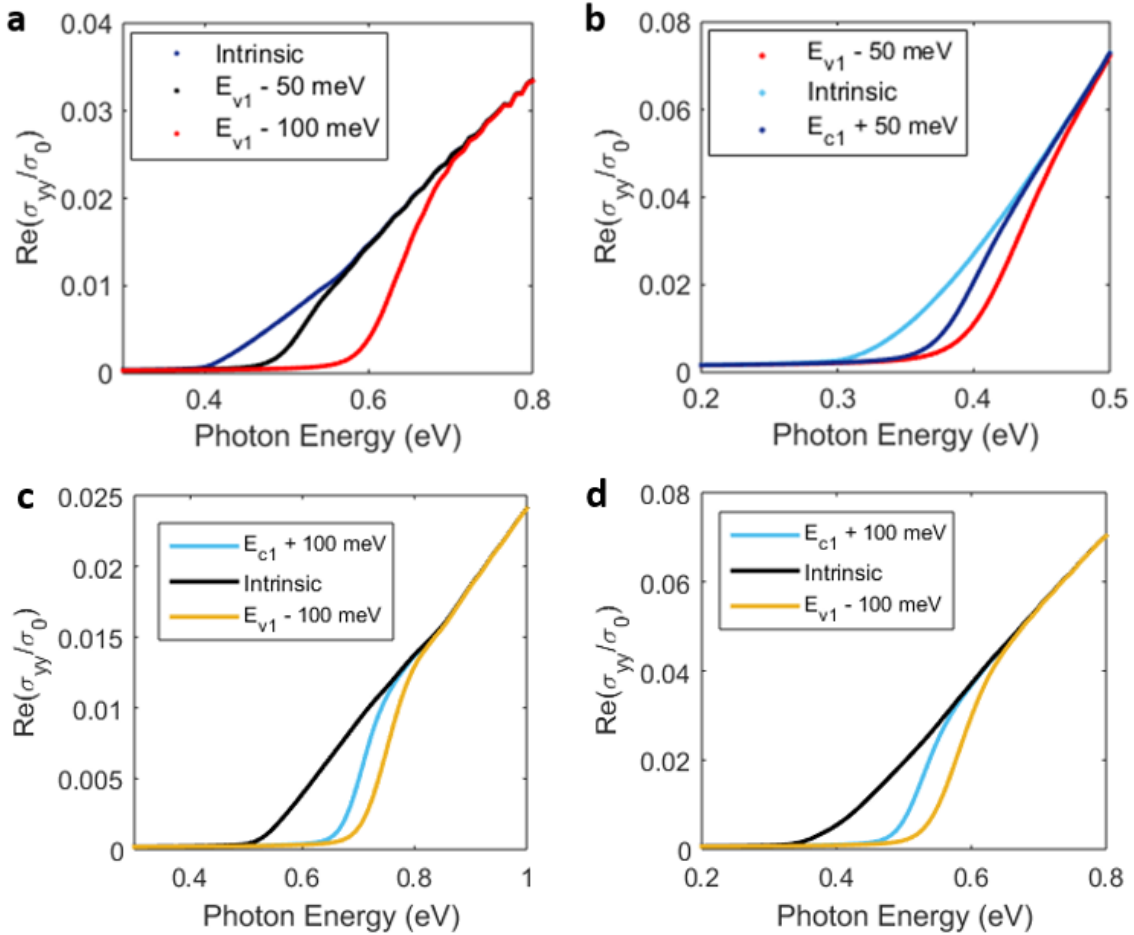

**Figure S4:** Calculated $\sigma_{yy}$ optical conductivities at different Fermi levels for **a)** the 6.5 nm flake, **b)** the 25 nm flake, **c)** the 7 nm flake and **d)** the 14 nm flake. In all cases, $\sigma_{yy}$ is one to two orders of magnitude smaller than $\sigma_{xx}$, implying that the interaction of the FTIR beam with the flake is dominated by the $\sigma_{xx}$. As a result, any polarization of the FTIR beam for measurements of the 6.5 nm and 25 nm flakes effectively scales the gate modulation as the strength of the interaction of the beam with the $\sigma_{xx}$ vs $\sigma_{yy}$ optical conductivity components changes.

## IV. Carrier Concentration Determination

We estimate carrier concentration in our flakes from gated resistance measurements by noting the applied bias at which the flake is approximately charge neutral – ie, least conductive – and using a parallel plate capacitor model to calculate the charge added between that bias and 0 V. For 285 nm silicon oxide, the parallel plate model results in a capacitance per unit area of $c = \varepsilon/d = 12$ nF/cm$^2$. We then calculate $\Delta Q = C\Delta V$.

## V. Accumulation/Depletion Length Determination

We estimate the screening length in our flakes using the Thomas-Fermi method adopted for black phosphorus by Tony Low, et al. The result of this calculation is shown in Figure S5, which describes band bending in the film as a function of depth / thickness. The screening length is of order 3 nm, indicating that band bending yields modulation that varies significantly along the depth axis of our flakes.

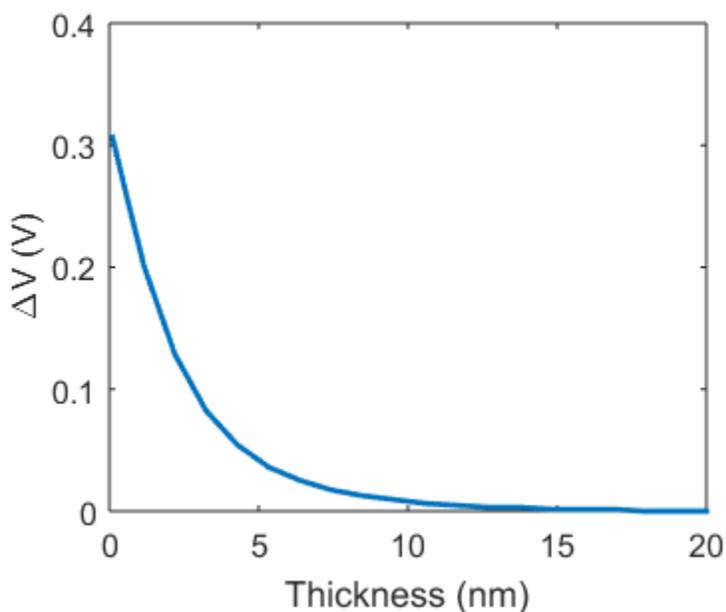

**Figure S5:** Calculated approximate voltage drop across the flake using the Thomas-Fermi method. The screening length is of order 3 nm.

## VI. Thickness Characterization

Atomic Force Microscopy (AFM) was used to determine the nominal thickness of the Black Phosphorus samples analyzed. Scans are shown in Figure S6.

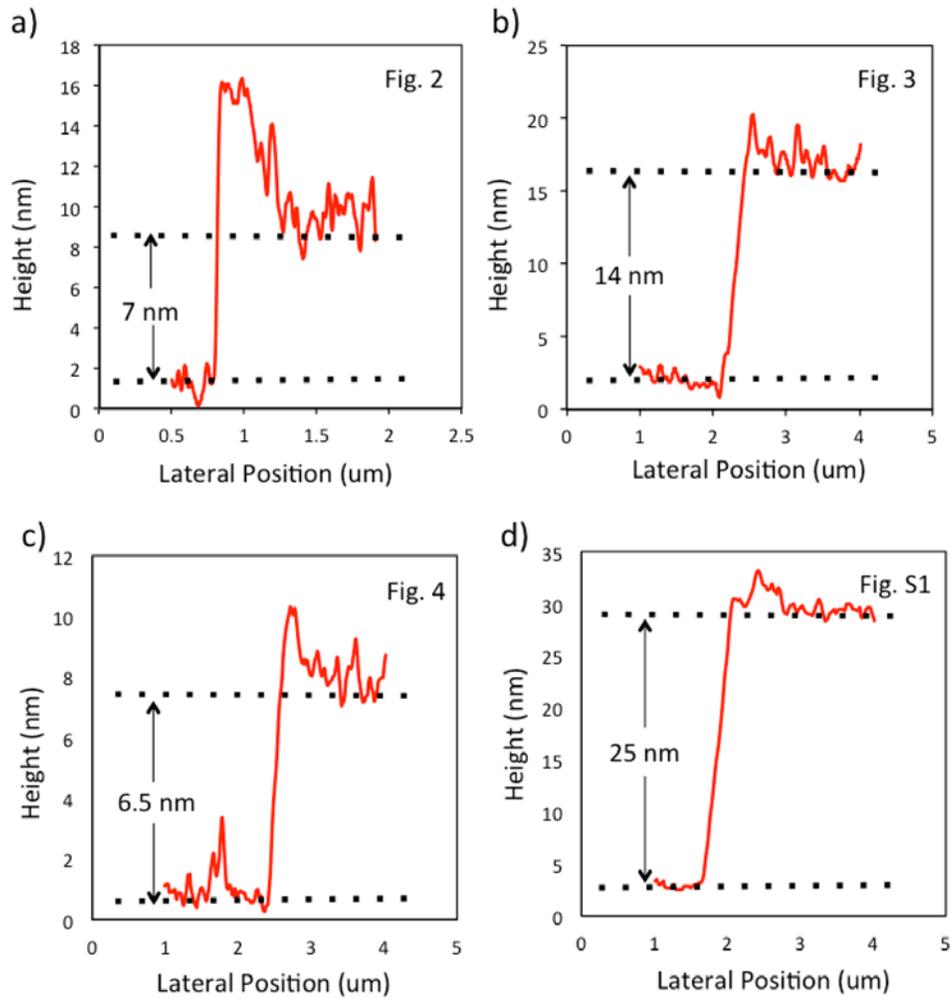

**Figure S6:** AFM Scans of BP samples presented in the main text and Supporting Information. a) Fig. 2, indicating 7 nm thickness b) Fig. 3, indicating 14 nm thickness, c) Fig. 4, indicating 6.5 nm thickness, d) Figure S1, indicating 25 nm thickness.

## VII. Raman Characterization

Raman spectroscopy was used to both compare results to standard literature for black phosphorus, and to characterize oxidation as a function of time. No appreciable change are seen in our encapsulated devices, suggesting that while oxide appears to form at some point during fabrication, it does not continue to form during our measurements.

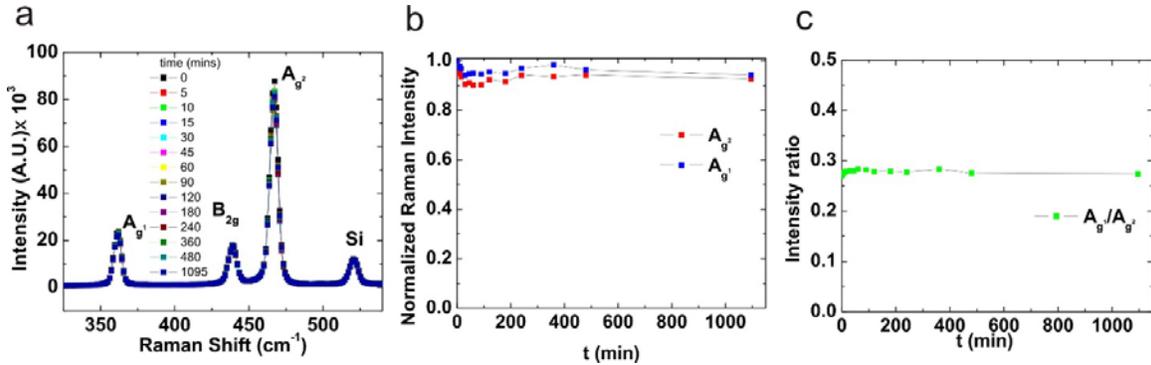

**Figure S7: a)** Raman spectrum of a few layer BP flake using a 514 nm laser showing the $A_g^1$, $B_{2g}$ and $A_g^2$ peaks. The $A_g^2$ peak position suggests a flake thickness of 4-6 nm. Spectra acquired at times ranging from 0 to 1095 mins are overlayed. No appreciable shifts in peak in changes in peak magnitude are visible as a function of time. **b)** Normalized intensity of $A_g^1$ and $A_g^2$ peaks as a function of time again suggesting no signs of oxidation degradation or decay. **c)** Time evolution of ratio of $A_g^1/A_g^2$ (known to be a clear indicator of oxidation) suggesting no appreciable change over 1095 mins further indication of no appreciable oxidation or degradation outside of the initial/immediate oxide formation upon exfoliation.